\documentclass[11pt]{iopart}
\usepackage{graphicx}
\usepackage{epsf,amsfonts,amssymb,amsbsy,wrapfig,cite}
\begin{document}
\title[Radial geodesics as a microscopic origin of black hole entropy]
{Radial geodesics as a microscopic origin of black hole entropy.\\
III: Exercise with the Kerr-Newman black hole}
\author{V.V.Kiselev*\dag}
\address{*
\ Russian State Research Center ``Institute for
High Energy
Physics'', 
Pobeda 1, Protvino, Moscow Region, 142281, Russia}
\address{\dag\ 
Moscow
Institute of Physics and Technology, Institutskii per. 9,
Dolgoprudnyi Moscow Region, 141700, Russia}
\ead{kiselev@th1.ihep.su}
\begin{abstract}
We specify an angular motion on geodesics to reduce the problem to
the case of radial motion elaborated in previous chapters. An
appropriate value of entropy for a charged and rotating black hole
is obtained by calculating the partition function on thermal
geodesics confined under horizons. The quantum aggregation is
classified in a similar way to the Reissner--Nordstr\o m black
hole.
\end{abstract}


\section{Continuation of \textit{Preface}}

A rotation of Kerr--Newman black hole involves a new feature in
the description of geodesics responsible for the entropy of black
hole: a nonzero projection of angular momentum on the axis of
rotation is permitted by the symmetry of the problem. Therefore,
we start with a specification of angular motion on geodesics of
conserved orbital momentum in Section 2. Then, the procedure has a
little to differ from the Reissner--Nordstr\o m black hole. The
difference is reduced to a particular dependence of mass sum on
the polar angle, that allows us to evaluate the partition function
and entropy for a cool state of aggregation in agreement with the
Bekenstein--Hawking formula \cite{Bekenstein,Hawking,Hawking2} in
Section 3. A short summary is situated in Section 4.

\section{Angular motion and reduced geodesics}
The Kerr--Newman metric of charged and rotating black hole can be
written in the form
\begin{equation}\label{KN}
    {\rm d}s^2_{\rm KN}=\frac{\Delta}{\Sigma}\,{\rm
    d}\omega_t^2-\frac{\Sigma}{\Delta}{\rm d}r^2-\Sigma\, {\rm
    d}\theta^2-\frac{\sin^2\theta}{\Sigma}\,{\rm d}\omega_\phi^2,
\end{equation}
with
\begin{equation}\label{defw}
    \begin{array}{rcl}
      {\rm d}\omega_t & = & {\rm d}t-a\,\sin^2\theta\,{\rm d}\phi,
      \\[2mm]
      {\rm d}\omega_\phi & = & (r^2+a^2)\,{\rm d}\phi-a\,{\rm d}t, \\
    \end{array}
    \qquad
    \begin{array}{rcl}
      \Sigma & = & r^2+a^2\,\cos^2\theta,
      \\[2mm]
      \Delta & = & (r-r_+)(r-r_-), \\
    \end{array}
\end{equation}
where the black hole parameters: mass $M$, charge $Q$, and angular
momentum $J$, are given by
\begin{equation}\label{MQJ}
    M=\frac{1}{2}(r_+ +r_-),\quad
    Q^2+a^2= r_+ r_-,\quad
    J =a\,M.
\end{equation}

In order to proceed with the Hamilton--Jacobi equation for a test
particle with a mass $m$
\begin{equation}\label{1}
    g^{\mu\nu}\;{\partial_{\mu} S_{HJ}}\,{\partial_{\nu} S_{HJ}} - m^{2} =
    0,
\end{equation}
one has to invert the metric in terms of $\{t,\phi\}$ given by the
following elements:
\begin{equation}\label{metr}
    \begin{array}{rcl}
      g_{tt} & = & \displaystyle
      \frac{1}{\Sigma}(\Delta-a^2\,\sin^2\theta),
      \\[2mm]
      g_{t\phi} & = & \displaystyle
      -\frac{a\,\sin^2\theta}{\Sigma}(\Delta-r^2-a^2),
      \\[2mm]
      g_{\phi\phi} & = & \displaystyle
      \frac{\sin^2\theta}{\Sigma}(a^2\,\sin^2\theta-r^2-a^2).
      \\[2mm]
    \end{array}
\end{equation}
For the corresponding $2\times 2$-matrix $\hat g$ we get the
determinant
\begin{equation}\label{det}
    {\rm det}\hat g = \frac{\Delta\,\sin^2\theta}{\Sigma^2}\,
    [-a^4\,\sin^4\theta-(r^2+a^2)^2+2(r^2+a^2)a^2\,\sin^2\theta],
\end{equation}
which enters the inverse matrix elements
\begin{equation}\label{invmetr}
      g^{tt} =\frac{1}{{\rm det}\hat g}\,g_{\phi\phi},\quad
      g^{t\phi} =-\frac{1}{{\rm det}\hat g}\,g_{t\phi},\quad
      g^{\phi\phi} =\frac{1}{{\rm det}\hat g}\,g_{tt}.
\end{equation}
Then, we introduce two integrals of motion: an energy ${\cal E}$
and an angular momentum $\mu$, which determine the action in the
form
\begin{equation}\label{2}
    S_{HJ} = -{\cal E}\, t+\mu\,\phi+{\cal S}_{HJ}(r),
\end{equation}
at $\dot\theta\equiv 0$. From (\ref{1}) we deduce
\begin{equation}\label{3}
    \left(\frac{\partial {\cal S}_{HJ}}{\partial r}\right)^2
    =\frac{\Delta}{\Sigma}\left[{\cal E}^2\,g^{tt}+2{\cal E}\mu\,g^{t\phi}+\mu^2\,
    g^{\phi\phi}-m^2\right]\equiv \frac{\Delta}{\Sigma}\,h.
\end{equation}
which results in
\begin{equation}\label{4}
    {\cal S}_{HJ}(r) = \int \limits_{r_0}^{r(t)} \textrm{d}r\;
    \sqrt{\frac{h\,\Delta}{\Sigma}}.
\end{equation}
The trajectory is implicitly determined by equations
\begin{eqnarray}
  \frac{\partial S_{HJ}}{\partial {\cal E}} &=& t_0 = -t
  +\int\limits_{r_0}^{r(t)} \textrm{d}r\; \sqrt{\frac{\Delta}{h\,\Sigma}}\,
   [{\cal E}\,g^{tt}+\mu\,g^{t\phi}], \label{p1}
  \\
  \frac{\partial S_{HJ}}{\partial \mu} &=& \phi_0 =
  \phi +\int\limits_{r_0}^{r(t)} \textrm{d}r\;\sqrt{\frac{\Delta}{h\,\Sigma}}\,
  [{\cal E}\,g^{t\phi}+\mu\,g^{\phi\phi}].\label{p2}
\end{eqnarray}
Taking the derivative of (\ref{p1}), (\ref{p2}) with respect to
the time\footnote{As usual $\partial_t f(t) =\dot f$.}, we get
\begin{eqnarray}
  1 &=& \dot r\; \sqrt{\frac{\Delta}{h\,\Sigma}}\,
   [{\cal E}\,g^{tt}+\mu\,g^{t\phi}], \\
  \dot \phi &=& -\dot r\; \sqrt{\frac{\Delta}{h\,\Sigma}}\,
  [{\cal E}\,g^{t\phi}+\mu\,g^{\phi\phi}],
\end{eqnarray}
determining the angular motion by
\begin{equation}\label{dotp}
    \dot\phi=-\frac{{\cal E}\,g^{t\phi}+\mu\,g^{\phi\phi}}
    {{\cal E}\,g^{tt}+\mu\,g^{t\phi}}.
\end{equation}
Further, we use a relation specifying the angular motion by
\begin{equation}\label{spec}
    \mu ={\cal E}\,a\,\sin^2\theta,
\end{equation}
that gives
\begin{equation}\label{dotp2}
    \dot\phi =\frac{a}{r^2+a^2}.
\end{equation}
Note, for such the angular velocity ${\rm d}\omega_\phi=0$, and
the interval takes the form
\begin{equation}\label{ds}
    {\rm d}s^2 =\Sigma\,\frac{\Delta}{(r^2+a^2)^2}\big({\rm d}t^2-{\rm
    d}r_*^2),
\end{equation}
with
\begin{equation}\label{dif}
    {\rm d}r_* =\frac{r^2+a^2}{\Delta}\,{\rm d}r,
\end{equation}
yielding
\begin{equation}\label{r}
    r_*=r+\frac{r_+^2+a^2}{r_+-r_-}\,\ln\left[\frac{r}{r_+}-1\right]-
    \frac{r_-^2+a^2}{r_+-r_-}\,\ln\left[\frac{r}{r_-}-1\right].
\end{equation}
Then, we can repeat the Hamilton--Jacobi formalism for the
interval of reduced motion in (\ref{ds}) and find
\begin{equation}\label{HJ}
    \frac{1}{m^2}\,\left(\frac{\partial {\cal S}_{HJ}}{\partial
    {r_*}}\right)^2={\cal E}_A-U(r),
\end{equation}
with ${\cal E}_A=1/A$,
\begin{equation}\label{U}
    U(r) =\frac{\Sigma\cdot\Delta}{(r^2+a^2)^2},
\end{equation}
and
\begin{equation}\label{dott}
    \frac{{\rm d}t}{{\rm d}r_*}=\frac{{\cal E}_A}{\sqrt{{\cal
    E}_A-U(r)}}.
\end{equation}
Therefore, on the geodesics we get the causal interval
\begin{equation}\label{caus}
    {\rm d}s^2_A =\frac{U^2(r)}{{\cal
    E}_A-U(r)}\,\frac{(r^2+a^2)^2}{\Delta^2}\,{\rm d}r^2.
\end{equation}

For the ground state we will consider further, ${\cal E}_A\to -0$
and \begin{equation}\label{ground}
    {\rm d}s^2=\frac{r^2+a^2\,\cos^2\theta}{(r_+-r)(r-r_-)}\,{\rm
    d}r^2,\qquad r_-<r<r_+,
\end{equation}
while the increment of time per cycle is given by
\begin{equation}\label{inc}
    \Delta_c t_E=2\pi\,(r_++r_-).
\end{equation}

Further, due to (\ref{ds}) and (\ref{r}) we can introduce two
consistent maps under the horizons in terms of Kruskal isotropic
variables in a manner of our treatment in Chapter II, with inverse
temperatures
\begin{equation}\label{beta}
    \beta_+=4\pi\frac{r_+^2+a^2}{r_+-r_-}, \qquad
    \beta_-=4\pi\frac{r_-^2+a^2}{r_+-r_-},
\end{equation}
and a quantum ratio of horizon areas
\begin{equation}\label{area}
    \frac{{\cal A}_+}{{\cal A}_-}=\frac{4\pi(r_+^2+a^2)}{4\pi(r_-^2+a^2)}=l \in
    \mathbb N.
\end{equation}
The winding numbers on geodesics are equal to the same values as
in Chapter II,
\begin{equation}\label{wind}
    n_+=\frac{2 l}{l-1},\quad n_-=\frac{2}{l-1}.
\end{equation}

The increment of interval per cycle $\Delta_c s(\cos\theta)$
follows from (\ref{ground}), while at $\theta=\pi/2$ we easily get
\begin{equation}\label{/2}
    \Delta_c s(0) =\pi (r_++r_-).
\end{equation}
Thus, we can follow the analogy with the Reissner--Nordstr\o m
black hole.

\section{Entropy}

\begin{figure}[th]
  \begin{center}
  \setlength{\unitlength}{1mm}
  \begin{picture}(100,62)
  \put(5,0){\includegraphics[width=9cm]{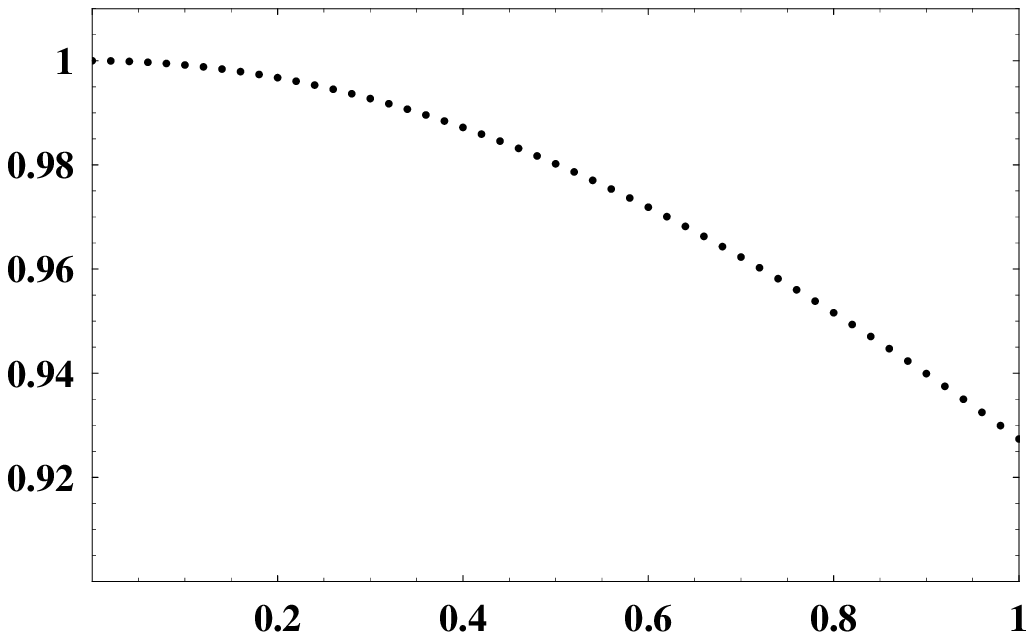}}
  \put(97,0){$\cos\theta$}
  \put(0,58){$\sigma_+(\cos\theta)$}
  \end{picture}
  \end{center}
  \caption{The variation of $\sigma_+$ versus $\cos\theta$ at $a=1$,
  $r_-=2$, $r_+=3$.}\label{mass}
\end{figure}

Introducing a sum of particles moving at a fixed value of angle
$\theta$ in the maps $\pm$,
\begin{equation}\label{sigma}
    \sigma_\pm(\cos\theta) =\sum_{\pm} mc\bigg|_\theta,
\end{equation}
for pure ``ice'' state of aggregation we get the partition
function
\begin{equation}\label{part+}
    \ln Z_+ =-n_+\,\Delta_c s(\cos\theta)\,\sigma_+(\cos\theta).
\end{equation}
In order to get the most probable configuration the product
$\Delta_c s(\cos\theta)\,\sigma_+(\cos\theta)$ should be invariant
under the variation of $\theta$ and take its minimal value, which
is reached at $\cos\theta=0$, so that
\begin{equation}\label{opt}
    \ln Z_+ =-n_+\,\Delta_c
    s(0)\,\sigma_+(0)=-\frac{\beta_+}{2}\,\sigma_+(0).
\end{equation}
Thus, in the thermal equilibrium, the sum of masses
$\sigma(\cos\theta)$ adjusts its value in order to compensate the
dependence of $\Delta_c s$ on $\cos\theta$. An example of such the
adjustment is shown in Fig. \ref{mass}.

From (\ref{opt}) and Chapter II we deduce the expressions
\begin{equation}\label{ss}
    \sigma_+(0)= 2M-\frac{1}{2\beta_+}\,{\cal A}_+,
\end{equation}
and the entropy
\begin{equation}\label{entr}
    {\cal S}_+=\frac{1}{4}\,{\cal A}_+,
\end{equation}
valid due to the standard relation between the temperature and
`surface gravity' (see discussion in review \cite{Damour}),
\begin{equation}\label{stand}
    T_+=4\,\frac{\partial M}{\partial {\cal A}_+},\qquad \mbox{at
    } {\rm d}Q^2\equiv 0,\quad {\rm d}J\equiv 0.
\end{equation}

A discrimination of two phases of aggregation is the same as for
the charged black hole.

\section{Conclusion}
In present chapter we have shown how the consideration of rotating
black hole is reduced to the charged one of Reissner and Nordstr\o
m. The only actual difference is the adjustment of mass sum versus
the polar angle.

Finally, let us make a short note on the extremal black hole with
$r_+=r_-$, that corresponds to solitonic BPS states in
superstrings \cite{StromingerVafa+}. We have found that the
extremum takes place, when the sphere of euclidian time and radius
degenerates to a torus with an infinitely small radius (two poles
of sphere are glued, but the radius tends to zero). The
corresponding winding number tends to infinity for the ground
state. However, coming from such the singular ground state to an
excited state, one could adjust the level of winding number (the
value of $x$) and tune the sum of particle masses in order to
preserve the final value of entropy, which is independent of
particular value of temperature at the given area of external
horizon. Due to the thermal quantization of ratio of horizon
areas, the limit of extremal black hole cannot be reached
continuously: one has got a quantum jump of $r_-$.

Thus, we finalize considering the basics of method for the
evaluation of black hole entropy by calculating the microscopic
partition function on classical geodesics confined under the
horizons by their thermal quantization. We hope that the tool
provides a new insight to the entropy of black hole, not excluding
some new particular problems and questions.

\vspace*{3mm}
 This work is partially supported by the grant of the
president of Russian Federation for scientific schools
NSc-1303.2003.2, and the Russian Foundation for Basic Research,
grants 04-02-17530, 05-02-16315.


\end{document}